\documentclass[12pt,preprint]{aastex}

\def\plotfiddle#1#2#3#4#5#6#7{\centering \leavevmode
    \vbox to#2{\rule{0pt}{#2}}
    \includegraphics{#1}}

\newcommand{\Mstar}{M_*}
\newcommand{\Mdisk}{M_{\rm disk}}
\newcommand{\Msun}{M_{\odot}}

\newcommand{\percc}{\rm \,cm^{-3}}

\newcommand{\persqcm}{\rm \,cm^{-2}}

\newcommand{\pers}{\,{\rm s}^{-1}}

\newcommand{\ccps}{{\rm cm}^{3}\,{\rm s}^{-1}}

\newcommand{\Htwo}{{\rm H_{2}}}

\newcommand{\HtwoO}{{\rm H_2O}}
\newcommand{\CtwoHtwo}{{\rm C_2H_2}}

\newcommand{\COtwo}{{\rm CO_2}}

\def\micron{\hbox{$\,\mu$m}}

\def\micron{\hbox{$\,\mu$m}}
\newcommand{\be}{\begin{equation}}
\newcommand{\ee}{\end{equation}}

\begin{document}


\title{The HCN-Water Ratio in the Planet Formation Region of Disks } 


\author{Joan R.\ Najita}
\affil{National Optical Astronomy Observatory, 950 N. Cherry Avenue, Tucson, AZ, 85719}

\author{John S.\ Carr}
\affil{Naval Research Laboratory, Code 7211, Washington, DC 20375}
\author{Klaus M.\ Pontoppidan}
\affil{Space Telescope Science Institute, 3700 San Martin Drive, Baltimore, MD 21218}
\author{Colette Salyk}
\affil{National Optical Astronomy Observatory, 950 N. Cherry Avenue, Tucson, AZ, 85719}
\author{Ewine F.\ van Dishoeck}
\affil{Leiden Observatory, Leiden University, P.O.\ Box 9513, 2300 RA Leiden, Netherlands}
\author{Geoffrey A.\ Blake}
\affil{Division of Geological \& Planetary Sciences, Mail Stop 150-21, 
California Institute of Technology, Pasadena, CA 91125}



\begin{abstract}
We find a trend between the mid-infrared HCN/$\HtwoO$ flux ratio 
and submillimeter disk mass among T Tauri stars in Taurus.
While it may seem puzzling that the molecular emission properties of 
the inner disk ($<$few AU) are related to the properties of the 
outer disk (beyond $\sim20$\,AU) probed by the submillimeter 
continuum, an interesting possible interpretation is 
that the trend is a result of planetesimal and protoplanet 
formation.
Because objects this large are decoupled from the accretion flow, 
when they form, they can lock up water (and oxygen) beyond 
the snow line, thereby enhancing the C/O ratio in the inner disk and 
altering the molecular abundances there. 
We discuss the assumptions that underlie this interpretation, 
a possible alternative explanation, and related open questions that 
motivate future work. 
Whatever its origin, understanding the meaning of the relation 
between the HCN/$\HtwoO$ ratio and disk mass is of interest as trends 
like this among T Tauri disk properties are relatively rare.
\end{abstract}






\section{Introduction}

Emission from gaseous water and organic molecules such as HCN are common 
in the mid-infrared spectra of T Tauri stars, as measured with the 
{\it Spitzer Space Telescope}
(Carr \& Najita 2011; Salyk et al.\ 2011; Pontoppidan et al.\ 2010; 
Pascucci et al.\ 2009).
These emission features are believed to arise from the warm atmosphere of 
the disk within a few AU of the star, based on estimates of the temperature and 
emitting area of the emission (Carr \& Najita 2008, 2011; Salyk et al.\ 2008, 2011). 
This interpretation is supported by recent thermal-chemical models 
of gaseous inner disk atmospheres, which can account for 
the temperatures, column densities, and emitting areas of 
many of the observed molecular species 
(Najita, \'Ad\'amkovics, \& Glassgold 2011; see also 
Ag\'undez et al.\ 2008; Woitke et al.\ 2009; Willacy \& Woods 2009; 
Heinzeller et al.\ 2011).  

While molecular emission is common, the observed spectra of T Tauri stars 
show significant diversity. 
As one example, the water emission strength may be larger or smaller 
than that of HCN, depending on the source. 
Carr \& Najita (2011) noted that the flux ratio of HCN to $\HtwoO$ 
emission hinted at an increasing trend with disk mass, based on 
their small sample of objects.  
They noted that such a trend could result if higher mass disks are 
more efficient in forming large icy bodies 
($>$\,km in size; e.g., planetesimals and protoplanets) and consequently 
in locking up water in the outer disk. Because this sequestration of 
water in the outer disk (beyond the snow line) would reduce the 
water (and oxygen) abundance of the inner disk (e.g., Ciesla \& Cuzzi 2006), 
the inner disks of such systems would have an enhanced C/O ratio and, 
consequently, higher abundances of organic molecules. 
Najita et al.\ (2011) showed that only modest enhancements in the 
C/O ratio of the inner disk 
atmosphere are needed to produce a significant increase in the relative 
column densities of warm HCN and $\HtwoO$ in the disk atmosphere. 

Finding possible evidence of planetesimal and protoplanet formation, 
even if indirect, is of interest as a test of planet formation 
theory, as planetesimals and protoplanets are the building blocks 
of ice and gas giant planets in core accretion theory. 
Currently, debris disks provide perhaps the best evidence for planetesimal 
formation outside the solar system (P.\ Armitage, personal communication).   
The many debris disks that are now known suggest that
planetesimals form efficiently and commonly, even though their 
formation mechanism remains uncertain (see Chiang \& Youdin 2010 for
a review of our understanding of planetesimal formation).  
The high occurence rate of extrasolar giant planets among 
mature solar-type stars 
further suggests that planetesimals and protoplanets must form commonly and 
within a Myr if core accretion is to operate within the few Myr gas dissipation 
timescale of disks (Hillenbrand 2008 and references therein). 

One way to test this hypothesis is to look 
for evidence of planetesimal and protoplanet formation in Myr old 
protoplanetary disks.
Here we take a step in this direction, by exploring in a larger sample 
the possible trend reported by Carr \& Najita (2011) involving the 
ratio of HCN/$\HtwoO$ emission strength in T Tauri disks.

\section{Observations}

We obtained spectra of 18 classical T Tauri stars 
with {\it Spitzer}/IRS (Houck et al.\ 2004) as part of a GO-5 program 
(PID 50641, PI J.\ Carr). 
The results for one of the sources (DR Tau) has been discussed 
in previous reports on this program (Pontoppidan et al.\ 2010; 
Salyk et al.\ 2011).
Because new observations in the IRS short-high module were not obtained
for one of the GO-5 targets (IQ Tau), data for this star were obtained
from the {\it Spitzer} archive (from PID 172, Evans et al.\  2003; 
see Pontoppidan et al.\ 2010).
These data were combined with data for 9 additional Taurus 
targets that were obtained in a previously reported 
GO-2 program (PID 2300; Carr \& Najita 2011, 2008) 
for a total sample of 28 objects.  

All of the sources reported here 
are located in the Taurus-Auriga star forming region.  
We focused on the Taurus sources from among the GO-5 targets, 
because they have submillimeter disk masses that 
have been estimated in a homogeneous way (Andrews \& Williams 2005). 
Because the targets are all well-known members of Taurus, we can 
assume that the sample is a coeval population that originated 
from cloud cores of similar metallicity. 

Our sample (Table 1) was chosen to span a broad range of stellar 
accretion rates, X-ray luminosities, and 
mid-infrared colors, but only a limited range in spectral type 
(primarily K3--M1). 
To characterize mid-infrared color, we used the spectral index 
between $13\micron$ and $25\micron$, $n_{13-25}$, as defined and 
reported by Furlan et al.\ (2006).
The sample excludes T Tauri stars with the extreme mid-infrared 
colors of transition objects (roughly $n_{13-25}>1.5$; Furlan et al.\ 2006). 
Most sources are contained within the mid-infrared color bounds of 
``normal T Tauri stars'' (roughly $n_{13-25} < 0.5$) 
as denoted by Furlan et al.\ (2006), although a few have 
intermediate mid-infrared colors 
(Haro 6-13, HK Tau, and SU Aur; $n_{13-25}$ of 0.5--1.5).
The submillimeter disk masses of the sources also span a broad range.  
Table 1 gives the disk masses derived by 
Andrews \& Williams (2005), primarily  
from spectral energy distribution (SED) fits, assuming a gas-to-dust mass ratio of 100.

None of the sources is known to have a companion close enough to 
dynamically alter the inner few AU of the disk, the region from 
which the {\it Spitzer} molecular emission arises (Carr \& Najita 2008, 2011; 
Salyk et al.\ 2011). 
However, several of the sources have companions close enough to be 
included in the $4.7\arcsec$-wide slit of IRS. 
Carr \& Najita (2011) discuss the companion properties of the GO-2 
sources DK Tau, RW Aur, and UY Aur. 
The other targets with nearby companions 
(GK Tau at 2.5$\arcsec$, HK Tau at 2.3$\arcsec$, 
and HN Tau at 3.1$\arcsec$) 
have flux ratios that are large, 
so the companion is not expected to contribute significantly 
to the {\it Spitzer} spectrum. 
The companions to HK Tau and HN Tau are, respectively, 
$\sim 30$ times fainter at $11.8\micron$ (McCabe et al.\ 2003) and 
$\sim 65$ times fainter at $L$ (White \& Ghez et al.\ 2001). 
The faint optical companion to GK Tau (e.g., Hartigan et al.\ 1994), 
which may be physically unrelated (White \& Ghez 2001; Itoh et al.\ 2008), 
is $\sim 60$ times fainter than GK Tau at $H$ (Itoh et al.\ 2008)
and is not detected with high-resolution MIR imaging (Skemer et al.\ 2011).

The spectra presented here were obtained with 
{\it Spitzer} IRS using the short-high (SH; 10--19$\micron$) module, 
which has a nominal resolving power of $\sim 600$.  
The spectra were obtained in staring mode, in which the target is 
placed at two nod positions along the length of the $11\arcsec$ by  
$4.7\arcsec$ slit. 
High signal-to-noise spectra were obtained through the use of a 
large number of nod cycles.  The resulting redundancy allowed us to 
evaluate the noise statistics and identify badly behaved pixels. 
Details of the data reduction procedure are given in Carr \& Najita (2011).
Table 1 gives a representative value of the signal-to-noise of 
the resulting spectra. 
In obtaining dedicated background observations for all targets,  
we avoided positions with continuum sources and known nebulosity. 
In most cases, the background integration time was one-half of 
the time spent observing the target.  

Here we report on the relation between HCN and $\HtwoO$ emission 
and disk mass.
We will provide a more in-depth description of the results for all of
the detected features in the SH and long-high (LH; 19--27$\micron$) modules in
in a subsequent report.

\section{Results}

\subsection{HCN and $\HtwoO$ Emission Strengths}

As found in previous studies, molecular emission is common in our sample 
and the emission spectra are rich. 
In particular, emission features of HCN and $\HtwoO$ are commonly detected (Fig.~1).    
Among the many water emission features present, the $17\micron$ group 
of features ($17.12\micron$, $17.22\micron$, $17.36\micron$) 
is an attractive measure of the water emission 
strength because these features are commonly detected, bright, 
and easy to measure, i.e., they are not blended with emission from 
other atoms or molecules. 
These features have been previously used to characterize the water emission 
from disks (Carr \& Najita 2011; Pontoppidan et al.\ 2010). 
Each feature is composed of multiple $\HtwoO$ transitions 
(e.g., Fig.~4 of Pontoppidan et al.\ 2010). 

To measure the strength of the HCN and $\HtwoO$ emission features, we
used synthetic slab models of the molecular emission from 
disks that include $\HtwoO$, OH, $\COtwo$, HCN, and $\CtwoHtwo$ 
(Carr \& Najita 2011; 
Carr \& Najita 2008) to 
identify continuum regions that are free of strong molecular 
emission. 
We identified continuum regions that bracket the HCN and $\HtwoO$ 
features of interest, as indicated in Figure 1. We then fit 
a linear continuum which we subtracted before 
measuring the flux of the water and HCN features.
Because of the possibility that unresolved spectral features 
are present in either the feature or continuum regions, we interpret 
the measured values as emission strengths that are more akin to 
spectral indices than true fluxes.

At {\it Spitzer} resolution, the HCN feature may be blended with 
water emission (see Carr \& Najita 2011, figure 4).
To estimate the significance of the water contamination, we can use 
the fluxes of other bright water lines in the spectrum as a predictor 
of the strength of the water emission in the HCN region.   
We examined 
synthetic slab models of water emission (e.g., as described in 
Carr \& Najita 2011) as a guide to which water features best 
track the water emission strength in the HCN band. 
We adopted the estimator
$0.4 (F_{15.17} + 0.8 F_{15.79}),$
where $F_{15.17}$ and $F_{15.79}$ are 
the emission strengths of the $15.17\micron$ and $15.79\micron$ water features,  
which predicts the contaminating water emission to within $\pm20$\% over 
the range of temperatures (500--700\,K) and 
column densities ($\log [N_\HtwoO/\persqcm] = 17.5-18.5$) found for 
water emission from T Tauri disks 
(Carr \& Najita 2011; Salyk et al.\ 2011). 

To estimate the strength of the HCN emission alone, we subtracted the 
estimated water emission in the HCN region using the above relation.  
The results discussed below are not sensitive to how this 
correction is made or whether it is carried out at all, as discussed 
in the following section.\footnote{A similar, but not identical, 
method of measuring the HCN and $\HtwoO$ 
fluxes was used in Carr \& Najita (2011). 
The $\HtwoO$ fluxes reported there are the combined 
flux of same the three $\HtwoO$ features measured here.
However, in the earlier study, the $\HtwoO$ flux was measured
in the entire spectral swath between $17.075\micron$ and $17.385\micron$.
The HCN flux was measured over the same spectral region, 
between $13.9\micron$ and $14.05\micron$, but after subtracting 
the best-fit slab emission models for $\HtwoO$, OH, and $\CtwoHtwo$,
a more time-consuming approach.
These details are unimportant to the results discussed here.}
The predicted strength of the $\HtwoO$ emission in 
the HCN region is typically $\sim 20$\% of the measured HCN emission.  

The emission strengths of HCN (before and after correction) and 
$\HtwoO$ are given in Table 1. The tabulated uncertainties 
only account for the contribution from the noise in the spectrum, 
whereas the use of a linear fit to the underlying continuum contributes 
additional uncertainty. For example, the inability to account for dust 
emission features with our simple method can lead to negative 
emission strengths in some cases 
(e.g., FN Tau, HQ Tau). 

Figure 2, which compares the $\HtwoO$ emission strength against the 
estimated HCN emission strength including the correction for 
contaminating water emission, shows that these are well correlated.  
The Kendall rank correlation test (28 sources) has a $\tau$ of 0.57 and 0.69 with 
and without the correction for contaminating water emission, respectively.
The associated two-sided P-values are $\sim 2\times 10^{-5}$ or smaller. 
The higher correlation value for uncorrected HCN emission strength 
probably results from water emission in the HCN region which is correlated 
with the $17\micron$ water emission that we measure. 
A related result, the correlation of HCN peak/continuum and $\HtwoO$ 
$17.2\micron$ peak/continuum, was shown in Salyk et al.\ (2011).

We might expect a correlation between HCN and $\HtwoO$ emission
from a disk atmosphere for several reasons.  Increased heating 
(mechanical or radiative) would tend to produce a warmer disk atmosphere 
(i.e., with a higher temperature and possibly a larger emitting area) 
or a deeper temperature inversion in the atmosphere, which 
would increase the emission from all molecules that are present in it.  
The removal of grains from the atmosphere through settling can also
increase the column density of the warm atmosphere that can be seen
in emission. 
A lower abundance of grains renders a larger column density of the 
disk atmosphere optically thin in the continuum, thereby ``revealing'' 
its emission.
In addition, if the heating of the disk atmosphere has significant 
contributions from processes other than (grain-related) photoelectric 
heating, the removal of grains also reduces gas-grain cooling and 
warms the atmosphere.  
These effects would also increase the emission from all molecules that 
are present in the atmosphere.

Grain settling can, in principle, also alter the relative flux 
from molecular species that coexist in a disk atmosphere. 
As deeper layers of the disk atmosphere become visible as grains settle, 
not only is the visible column density of each species increased 
(as described above), but also, molecular level populations are 
thermalized at different depths as denser layers, deeper in the disk, 
are reached. 
Because species with higher critical densities are thermalized at 
greater depth, an apparent change in molecular flux ratios may 
arise as deeper regions of the disk are probed.

\subsection{HCN/$\HtwoO$ Ratio and Disk Mass}

Although neither HCN (corrected for $\HtwoO$ contamination or not) 
nor $\HtwoO$ shows a trend individually with disk mass (Figure 4), 
the ratio of these quantities does show a trend with disk mass. 
Figure 3 plots the ratio of the (corrected) HCN to $\HtwoO$ emission 
strengths against disk mass. 
Sources with $\HtwoO$ emission strengths that are negative (HQ Tau) 
or near zero (FN Tau) were excluded from the plot. 
To highlight the properties of typical T Tauri stars, 
sources with a strong jet (DG Tau) or mid-infrared colors outside the 
``Normal T Tauri'' region of Furlan et al.\ (2006; $n_{13-25} > 0.5$) 
are not marked in red.  
The ratio of HCN/$\HtwoO$ emission strength increases by 
at least an order of magnitude over a range of $\sim 100$ in disk 
mass.  

The results confirm the possible trend between the HCN/$\HtwoO$ 
emission ratio and disk mass shown in Carr \& Najita (2011).
For the sources marked in red (17 sources), the Kendall rank $\tau$ 
is 0.61 with an associated two-sided P-value of $6\times 10^{-4}$. 
A similar trend is found between the HCN/$\HtwoO$ emission ratio 
and the $850\micron$ flux; 
this is expected because the 
$850\micron$ flux is a rough predictor of disk mass 
while a more precise measurement is obtained by modeling 
the SED over a wider range of wavelengths  
(Andrews \& Williams 2005).
The correction to the
HCN flux for contaminating water emission makes no significant difference
to this result; using the uncorrected HCN strengths gives
a Kendall rank $\tau$ of 0.58 and a two-sided P-value of 
$1 \times 10^{-3}$.  

One might wonder if the trend with disk mass really reflects a trend 
with stellar mass, as $\Mstar$ may be proportional to $\Mdisk$ in an 
average sense. This is implausible because our sample spans 
a narrow range of spectral types (Table 1) and hence stellar masses 
($\sim 0.5-1.2\Msun$, e.g., White \& Ghez 2001).
We also confirmed that the observed trend is not driven by mid-infrared color 
(cf.\ section 3.1). 
Subsamples with redder or bluer mid-infrared color separately show the same trend
between HCN/$\HtwoO$ and disk mass seen in Figure 3.

Given the trend between disk mass and the HCN to $\HtwoO$ 
emission ratio, it is worth considering what these quantities 
measure.  Within the context of homogeneous LTE slab models of these features
(Salyk et al.\ 2011; Carr \& Najita 2011),
the molecular fluxes depend on the column density, temperature, and
emitting area for each molecule.
For $\HtwoO$, the IRS spectrum contains both optically thick and optically 
thin lines over a range of excitation energies; as a result, we can characterize 
the typical temperature, column density, and area of the emitting region. 
The LTE slab modeling finds similar $\HtwoO$ column densities and
temperatures for the sources studied. The range in $\HtwoO$ emission strength
is therefore primarily a function of the $\HtwoO$ emitting area. 
In such a model, the total $\HtwoO$ mass contributing to the emission is the 
column density times the emitting area. Thus, to first order, the $\HtwoO$ 
flux provides a measure of the emitting mass of $\HtwoO$.

For HCN, there is some degeneracy between the column density and area 
of the emitting region.
However, because the emission is optically thin (or only marginally optically 
thick),
the HCN emission flux is roughly proportional to the HCN emitting mass.
Therefore, the LTE slab modeling suggests that the HCN/$\HtwoO$ flux ratio  
provides an
{\it approximate} 
measure of the relative emitting masses of HCN and $\HtwoO$ in the inner disk.

It is important to note, however, that sub-thermal collisional excitation
of both $\HtwoO$ and HCN is probable
(see Meijerink et al.\ 2009 for a discussion of non-LTE modeling of $\HtwoO$).
If there are differences in the critical densities of the transitions 
that make up the $\HtwoO$ and HCN features, then object to object differences 
in the density of the molecular emitting region
could induce variations in the HCN/$\HtwoO$ flux ratio.
If this is the case, then the HCN/$\HtwoO$ 
flux ratio could be more sensitive to the excitation conditions than 
to the relative amount of HCN and $\HtwoO$.
In section 4.2, we discuss whether this effect could produce
the observed trend in HCN/$\HtwoO$ flux with disk mass.

We might similarly consider what disk mass measures. 
As noted above, the disk mass plotted in Figure 3 is based on 
submillimeter dust emission, which 
is sensitive to the mass of dust in the outer disk (beyond $\sim 20$\,AU). 
In the next section we discuss why a property of the disk at such large radii 
might be related to a molecular emission property of the inner disk (within 
a few AU).

\section{Discussion}

We find that the possible trend of HCN/$\HtwoO$ emission strength
with disk mass shown in Carr \& Najita (2011) is confirmed
in a larger sample. The ratio of HCN/$\HtwoO$ emission strength
increases by at least an order of magnitude over a range of $\sim 100$ 
in disk mass.
It is perhaps surprising to find such a strong correlation between
a molecular emission property of the inner disk with a property of
the disk at much larger radii (the submillimeter dust mass).
This is more remarkable given the large number of processes that
could plausibly play a role in determining the molecular emission
strengths of inner disks (e.g., thermal, chemical, and photo-processes,
excitation).
There are few similarly clear trends between inner disk molecular
emission from T Tauri stars and other system parameters (Carr \&
Najita 2011; Salyk et al.\ 2011).  A similar kind of trend (e.g.,
of molecular ratios with disk mass) has not been reported in
millimeter studies of the outer regions of T Tauri disks (e.g.,
\"Oberg et al.\ 2011a).

\subsection{A Consequence of Planetesimal or Protoplanet Formation?}
We previously noted (Carr \& Najita 2011; Najita et al.\ 2011) that 
an increasing trend of HCN/$\HtwoO$ with disk mass 
might be expected to arise as a consequence of the formation of 
icy planetesimals and protoplanets. 
In the core accretion picture of planet formation, planets are 
built from collisions of planetesimals, solids larger than a 
kilometer in size, 
which themselves form from the aggregation of smaller solids. 
In the giant planet region of the disk 
(from the $\HtwoO$ snow line at a few AU to $\sim 20$\,AU),
the condensation of water ice onto grains removes a significant 
amount of oxygen from the gas phase; 
hence, the C/O ratio of the gas will be higher
than the bulk composition of the disk, and the C/O ratio
of the solids lower (see also \"Oberg et al.\ 2011b and 
their Figure 1).
As described by Ciesla \& Cuzzi (2006), 
as icy solids grow in this region of the disk (the outer disk), 
their migration, or lack thereof, can lead 
to gaseous inner disks that are either enhanced or depleted in 
water and oxygen. 
Resulting changes in the C/O ratio could 
affect the abundance of hydrocarbons in the inner disk 
(Najita et al.\ 2011).

More specifically, when icy grains in the outer disk are small
enough to couple well to the gas, water that is frozen on these
grains accretes along with the gas into the inner planet formation
region. When these grains pass the snow line and reach the inner disk, 
the water is returned to the gas phase as the ices evaporate, 
and the C/O ratio of the inner disk gas is representative
of the disk as a whole. 
However, when icy solids grow to larger 
(approximately meter) size, the strong headwind drag that they experience 
leads to rapid inward migration relative to the gas. 
When these bodies pass the snow line, their ices evaporate, 
hydrating the inner disk and,
in depositing water and oxygen, reduce the C/O ratio there. 
In contrast, 
when icy material in the outer disk reaches planetesimal or larger size, 
headwind drag becomes inconsequential, and these objects cease to migrate, 
sequestering water and oxygen beyond the snow line. 
As a result, the water- (and oxygen-) poor material that accretes 
into the inner disk enhances the C/O ratio there. 

These ideas suggest that {\it early} in the disk evolution process 
(before grain growth has occurred), the gaseous C/O ratio
of the inner disk will reflect the bulk C/O ratio
of the disk. At an {\it intermediate} stage, the inner disk will have a lower 
C/O ratio than the disk as a whole. At a {\it late} stage, the inner disk
will have a higher C/O ratio than the disk as a whole. 

In modeling the effect of the migration of solids on the evolution of the 
water distribution in protoplanetary disks, Ciesla \& Cuzzi (2006) 
found, in the specific scenarios they considered, that the 
gaseous water (and oxygen) abundance at AU distances could 
be enhanced or depleted by a factor of 10 in either direction
depending on the initial conditions and efficiency with which 
solids grow to large sizes. 
Such large variations would induce significant changes in the 
molecular abundances of the inner disk.

In their study of the thermal-chemical structure of inner disk 
atmospheres, Najita et al.\ (2011) looked at the 
effect of varying the inner disk C/O ratio on the column density 
of warm molecules in the disk atmosphere, i.e., the column that 
could produce the molecular emission observed with {\it Spitzer}.  
They found that 
increasing the C/O ratio of the inner disk from its nominal value of 
0.4 to 0.67 was enough to raise the ratio of the warm column of 
HCN to that of $\HtwoO$ by an order of magnitude,
similar to the range in the HCN/$\HtwoO$ emission ratio seen in 
Figure 3. The ratio increases dramatically because of both 
an increasing HCN abundance and a declining water abundance 
with increasing C/O. 
If the HCN/$\HtwoO$ emission ratios that we measure are 
related to their molecular abundances, these disk chemistry results 
suggest that variations in the C/O ratio 
of inner disks could plausibly explain the range of values we 
observe.  

What role could disk mass play in such a scenario to produce 
the trend we see in Taurus? The relation in Figure 3 could be 
explained if more massive 
disks have locked up a greater fraction of their icy solids in
non-migrating planetesimals and protoplanets, thereby 
increasing the C/O ratio of their inner disks.  
This would imply that a significant fraction of the systems in Taurus are in 
the ``late'' stage of planetesimal formation described above.
It seems reasonable that disks at the few Myr age of Taurus
would be in this stage, if a significant fraction of these systems
will go  on to form giant planets within the few Myr gas dissipation
timescale of disks.

More generally, in any given coeval star forming region, there may
be a spread in how far systems have evolved along the 
``early-intermediate-late'' sequence described above, 
with higher mass disks potentially
evolving more quickly through this sequence.
As a result, all star forming regions may not show the same trend seen 
here.  In a younger star forming region, the reverse may occur: if
the most massive disks are still in the ``intermediate'' stage and
less massive disks are in the ``early'' stage, then higher mass disks
would have inner disks with {\it lower} C/O ratios.

In applying the above picture of planetesimal formation to the 
interpretation of Figure 3, we are assuming that disk masses 
estimated from submillimeter measurements 
(which probe disk radii beyond 20\,AU) 
provide a useful indicator of the initial conditions in
the giant planet formation region (from the snow line to 20\,AU),
i.e., the region where icy planetesimals could form in sufficient 
abundance to affect the C/O ratio of the inner disk.   
Because grain evolution slows greatly with increasing disk radius 
(e.g., Dullemond \& Dominik 2005; Perez et al.\ 2012), it seems 
plausible that the {\it current} disk 
region beyond $\sim 20$\,AU would provide some insight into the mass 
that was {\it originally} present in the few--20\,AU region, 
at least in a relative sense. 
Our interpretation that the trend of HCN/$\HtwoO$ with
submillimeter disk mass reflects the enhanced 
formation of large icy bodies in more massive disks only assumes that 
the submillimeter disk mass gives a {\it relative} ordering 
of the initial mass in the planet formation region of the disk.
This perspective is similar to that adopted in other studies 
that assume (and argue why) submillimeter 
fluxes can be used to infer the planet-forming potential of a 
given disk (e.g., Greaves et al.\ 2007; Wyatt et al.\ 2007). 
 
This suggested interpretation highlights the 
possibility of looking for chemical signatures of observationally 
elusive steps in the planet formation process. 
A related idea has been explored in \"Oberg et al.\ (2011b)
regarding the effect of snow lines on the C/O ratio in planetary 
atmospheres.  They investigated how the different C/O ratios of the 
disk gas and solids as a function of disk radius can lead to different 
C/O ratios in the atmospheres of giant planets depending on how and 
where the planets acquired their atmospheres. 
As a result, the C/O ratio of the atmospheres of giant planets may provide 
a valuable chemical clue to that step in the planet formation 
process.

\subsection{Alternative explanation: subthermal excitation of HCN?}

In section 3 we described how the results of earlier LTE slab modeling 
suggest that the HCN/$\HtwoO$ flux ratio can probe the 
relative masses of HCN and $\HtwoO$ in the molecular emission region. 
It is important to examine the assumption of LTE, 
because non-LTE conditions might provide an 
alternative explanation for the trend we see. If HCN were subthermally
excited relative to $\HtwoO$, and if higher disk masses mean higher disk
densities overall, then the HCN/$\HtwoO$ emission ratio might increase
with disk mass simply because HCN is better excited, not because it 
is more abundant. 

The HCN-$\Htwo$ collisional rate coefficients 
at 300\,K (and higher) for the $\nu_2$ bending mode are 
$\sim 10^{-11}\ccps$ (Smith \& Warr 1991). 
For a typical Einstein A-value of $3\pers$, the HCN critical 
density is $\sim 2\times 10^{11}\percc$. 
In comparison, collisional rate coefficients for $\HtwoO$-$\Htwo$ 
correspond to critical densities of $10^9-10^{11}\percc$, 
depending on the transition. 
The densities in the warm atmosphere at the radii from
which the molecular emission is believed to arise 
($0.25-1$\,AU) are estimated to be $\sim 10^{10}-10^{12}\percc$ 
and vary in both vertical height and radius 
(e.g., in the models of Najita et al.\ 2011). 
Hence, HCN could be subthermally excited relative to $\HtwoO$ over 
some portion of the molecular emitting region. 
Detailed calculations are needed to understand whether 
and how strongly the HCN/$\HtwoO$ emission ratio is affected 
by the density of the disk atmosphere.

While density could play a role in the measured HCN/$\HtwoO$ emission
flux ratio, it is unclear whether this could produce a correlation
with disk mass. In order to produce the relation in Figure 3, the
submillimeter disk mass (which probes the region beyond 20\,AU)
would have to be a good indicator of {\it the density of the inner disk
atmosphere} (within a few AU) where the molecular emission arises.
The possibility of non-steady disk accretion, dead zones, and
accretion outbursts, suggests that the masses that currently reside
in the inner and outer disks might not be well correlated.\footnote{This
skepticism can be distinguished from our earlier assumption that
the outer disk (beyond 20 AU) might reasonably probe the {\it
initial} rather than {\it current} mass of the giant planet region
(few-20 AU).} 
Moreover, even if the inner disk mass is
correlated with the outer disk mass, the density in the molecular
line formation region of the upper atmosphere may not be as coupled.
As noted in section 3.2, the temperature and column density of the
warm disk atmosphere, as probed by the water emission, appears to
be similar among T Tauri disks (Carr \& Najita 2011; Salyk
et al.\ 2011).  This suggests that the density in the warm atmosphere
does not vary greatly.  We discuss this topic and other possibilities
for future work in the following section.

\section{Conclusions and Future Work}

We find a clear relation between the mid-infrared HCN/$\HtwoO$ 
flux ratio and submillimeter disk mass among T Tauri stars in Taurus.
It is perhaps surprising that emission properties of the gas in 
the inner disk ($<$ a few AU) are related to a property (disk mass)
probed by the dust in the outer disk (beyond $\sim 20$\,AU).

An interesting possible interpretation is that the trend arises 
as a consequence of the formation of icy planetesimals and 
protoplanets.
In this scenario, higher mass disks more readily form such large 
(non-migrating) bodies, which lock up water ice (and oxygen) beyond 
the snow line. 
The water- (and oxygen-) poor material 
that reaches the inner disk enhances the C/O ratio of the gas there. 
Because a higher C/O ratio both enhances the abundance of HCN and 
reduces the abundance of $\HtwoO$, the HCN/$\HtwoO$ ratio can 
rise dramatically as a result. 

This interpretation requires that 
{\it (i)} the HCN/$\HtwoO$ flux ratio is sensitive to 
the relative masses of HCN and $\HtwoO$ in the molecular emission region; 
{\it (ii)} submillimeter disk masses are a reasonable indicator of the 
initial masses in the icy planetesimal and protoplanet formation 
region of the disk (few to $\sim 20$\,AU); and 
{\it (iii)} higher mass disks in Taurus have formed non-migrating 
icy planetesimals more efficiently or more rapidly than lower mass disks.

An alternative explanation is that the variations in the HCN/$\HtwoO$
flux ratio are instead an excitation effect related to density.
This would require that the density of the warm molecular layer
of the inner disk where the emission arises is correlated with the
mass of the outer disk as measured by submillimeter dust emission.

We can explore these ideas in several ways. 
On the observational and modeling end, we need to measure relative 
molecular masses (rather than fluxes).
Currently, the errors on the molecular column density (or mass)
estimates from LTE analysis for HCN and $\HtwoO$ 
(Carr \& Najita 2011; Salyk et al.\ 2011) are too large to 
measure relative molecular masses. 
The higher spectral resolution ($R$=3000) of MIRI on {\it JWST} or
infrared echelle spectroscopy on future large telescopes, can
potentially resolve the modeling degeneracies that arise from the low-resolution
{\it Spitzer} spectra.

These measurements should be combined with a detailed non-LTE model for 
the excitation of HCN and $\HtwoO$.
With such a tool, one could also explore the counterhypothesis,
that the observed trend is a result of subthermal excitation of HCN
compared to $\HtwoO$, by looking for evidence of increased subthermal
excitation of HCN and $\HtwoO$ in sources with lower disk masses.
We should also search for trends involving other molecular ratios 
that would be affected by variations in the C/O ratio of the 
inner disk (possibly $\CtwoHtwo/\HtwoO$; e.g., Najita et al.\ 2011).  
On the theory end, we need a more complete model of the formation
of planetesimals and protoplanets, their efficiency of formation
as a function of disk mass, and their effect on the transport of
water in the disk, in order to properly evaluate whether our proposed
interpretation of the trend of HCN/$\HtwoO$ with disk mass is
tenable.



\acknowledgments
This work is based on observations made with the {\it Spitzer Space
Telescope}, which is operated by the Jet Propulsion Laboratory,
California Institute of Technology under a contract with NASA.
Support for this work was provided by NASA.
Basic research in infrared astrophysics at the Naval Research Laboratory 
is supported by 6.1 base funding. 
JN thanks Sharmila Dey for assistance with the data analysis.




\begin{center}
\begin{tabular}{lcrrrlr}    
\multicolumn{7}{c}{Table 1. Observations and Sample Properties} \\
\hline
Source    & s/n$^a$ &  F$_{\rm HCN}$ (uncorr.)$^b$ 
	 	    &  F$_{\rm HCN}$$^b$ \hfil\hfil 
		    &  F$_{\rm H_2O}$$^b$   \hfil\hfil 
		    & $\Mdisk/\Msun$ & $n_{13-25}$ \\
\hline
\hline
AA Tau     &  175  &  6.49  (0.12)  &  6.01  (0.13)  &  4.76  (0.19)  & 0.013  &  -0.37 \\
BP Tau     &  190  &  6.67  (0.14)  &  5.79  (0.14)  &  6.91  (0.21)  & 0.018  &  -0.24 \\
CI Tau     &  255  &  6.53  (0.14)  &  5.66  (0.14)  &  5.96  (0.20)  & 0.028  &  -0.02 \\
CW Tau     &  217  &  6.96  (0.31)  &  4.44  (0.32)  & 18.31  (0.46)  & 0.0024 &  -0.30 \\
CY Tau     &  145  &  1.42  (0.06)  &  1.24  (0.06)  &  1.00  (0.08)  & 0.006  &  -0.95 \\
DE Tau     &  310  &  1.46  (0.07)  &  1.24  (0.08)  &  1.55  (0.148) & 0.0052 &   0.12 \\
DG Tau     &  450  &  3.89  (0.73)  &  2.52  (0.78)  &  8.61  (1.35)  & 0.024  &   0.24 \\
DK Tau     &  222  &  7.94  (0.21)  &  5.97  (0.23)  & 15.10  (0.36)  & 0.005  &  -0.37 \\
DL Tau     &  347  &  5.22  (0.14)  &  4.78  (0.15)  &  2.52  (0.23)  & 0.09   &  -0.54 \\
DO Tau     &  345  &  4.22  (0.36)  &  3.55  (0.39)  &  4.51  (0.64)  & 0.007  &   0.08 \\
DP Tau     &  245  &  4.71  (0.18)  &  3.25  (0.19)  &  9.52  (0.31)  & $<$ 0.0005 &   0.02 \\
DR Tau     &  270  & 19.64  (0.42)  & 16.56  (0.45)  & 24.60  (0.79)  & 0.019  &  -0.27 \\
FN Tau     &  405  & -2.40  (0.12)  & -2.30  (0.13)  &  0.43  (0.23)  & ---    &   0.27 \\
FT Tau     &  245  &  2.28  (0.06)  &  1.97  (0.07)  &  2.21  (0.12)  & 0.014  &  -0.31 \\
FZ Tau     &  185  & 11.53  (0.36)  &  8.51  (0.37)  & 20.28  (0.40)  & 0.002  &  -0.56 \\
GI Tau     &  285  &  7.10  (0.18)  &  6.40  (0.19)  &  6.35  (0.26)  & ---    &  -0.40 \\
GK Tau     &  260  &  1.60  (0.19)  &  1.16  (0.21)  &  4.00  (0.38)  & 0.002  &   0.12 \\
Haro 6-13  &  295  &  1.76  (0.27)  &  0.73  (0.29)  &  7.45  (0.74)  & 0.011  &   0.90 \\
HK Tau     &  267  &  0.01  (0.05)  & -0.10  (0.06)  &  0.70  (0.13)  & 0.0045 &   1.12 \\
HN Tau     &  330  &  1.57  (0.17)  &  0.74  (0.18)  &  4.02  (0.35)  & 0.0008 &  -0.14 \\
HQ Tau     &  395  & -0.93  (0.16)  & -0.89  (0.17)  & -0.18  (0.31)  & 0.0005 &  -0.17 \\
IQ Tau$^c$ &  122  &  7.46  (0.22)  &  6.97  (0.22)  &  3.37  (0.32)  & 0.022  &  -0.90 \\
RW Aur     &  220  & 15.68  (0.45)  & 12.59  (0.46)  & 22.26  (0.63)  & 0.004  &  -0.27 \\
SU Aur     &  325  &  0.05  (0.45)  & -0.33  (0.48)  &  5.85  (1.43)  & 0.0009 &   1.26 \\
UY Aur     &  290  &  8.87  (0.56)  &  7.06  (0.59)  & 11.36  (1.10)  & 0.0018 &   0.17 \\
04216+2603 &  185  &  0.44  (0.07)  &  0.31  (0.07)  &  1.49  (0.11)  & ---    &  -0.33 \\
04303+2240 &  235  & 19.70  (0.47)  & 15.58  (0.50)  & 26.16  (0.69)  & ---    &  -0.74 \\
04385+2550 &  240  &  2.70  (0.12)  &  2.27  (0.13)  &  3.62  (0.22)  & ---    &   0.53 \\
\hline
\multicolumn{7}{l}{$^a$ Average of the continuum signal-to-noise at $14\micron$ and $17\micron$.}\\
\multicolumn{7}{p{1.0\textwidth}}{$^b$ Fluxes are in units of $10^{-17}{\rm W\,m^{-2}}$. Errors are 1-$\sigma$ 
uncertainties that only account for the noise in the spectrum.}\\
\multicolumn{7}{l}{$^c$ Archival data from the {\it Spitzer} c2d program (Evans et al.\ 2003).}\\
\end{tabular}
\end{center}
\clearpage

\begin{figure}
\plotfiddle{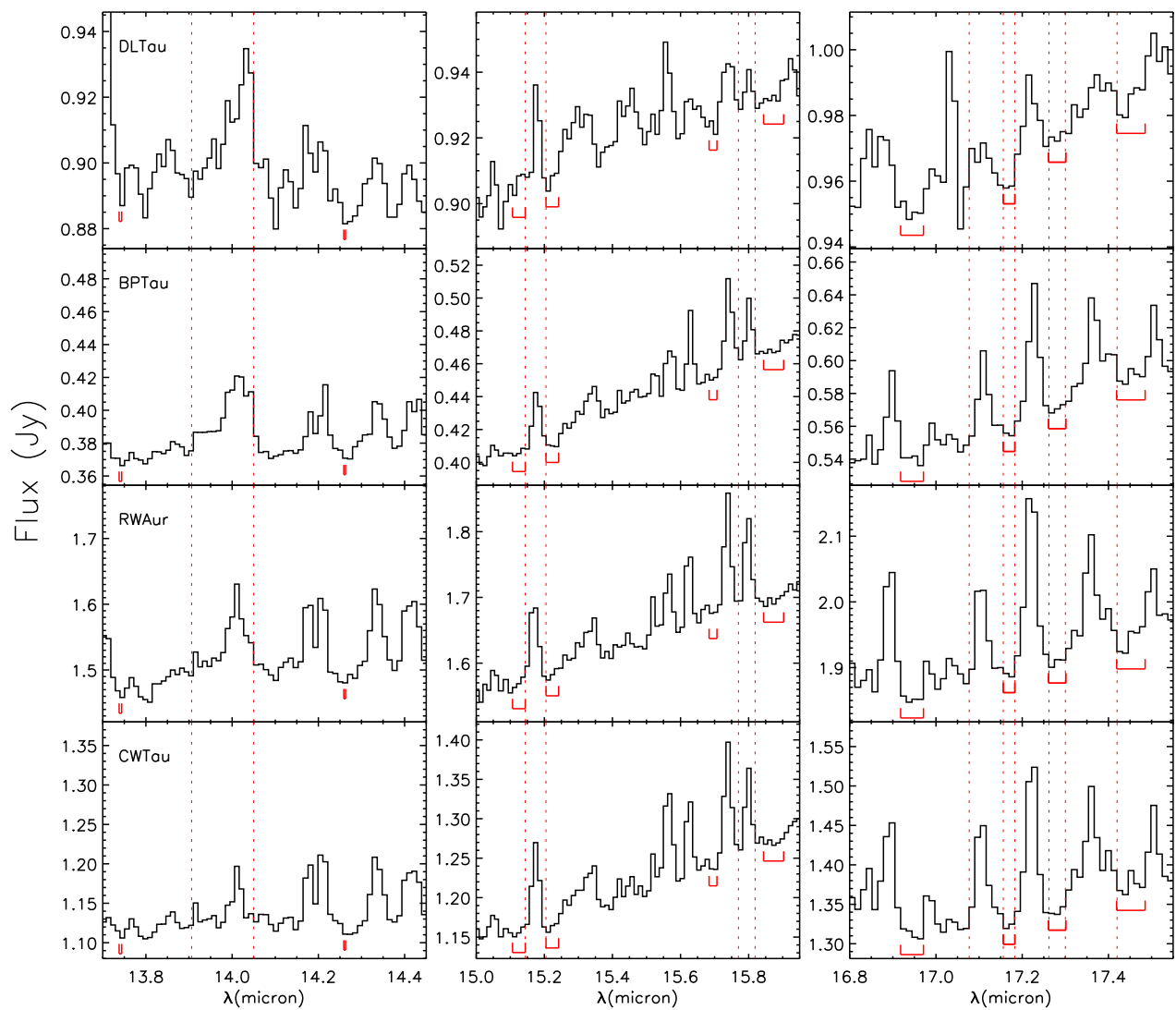}{6.0truein}{0}{100}{100}{-250}{0}
\caption{Representative spectra spanning a range of HCN vs.\ $\HtwoO$ 
emission strength.
The panels show HCN emission (left) and $\HtwoO$ emission features 
near $15\micron$ (center) and $17\micron$ (right).  
Vertical dotted lines indicate the spectral regions used to 
define the HCN and $\HtwoO$ features. 
Neighboring regions used to define the continuum 
are indicated by brackets below the spectrum. 
}
\end{figure}

\begin{figure}
\plotfiddle{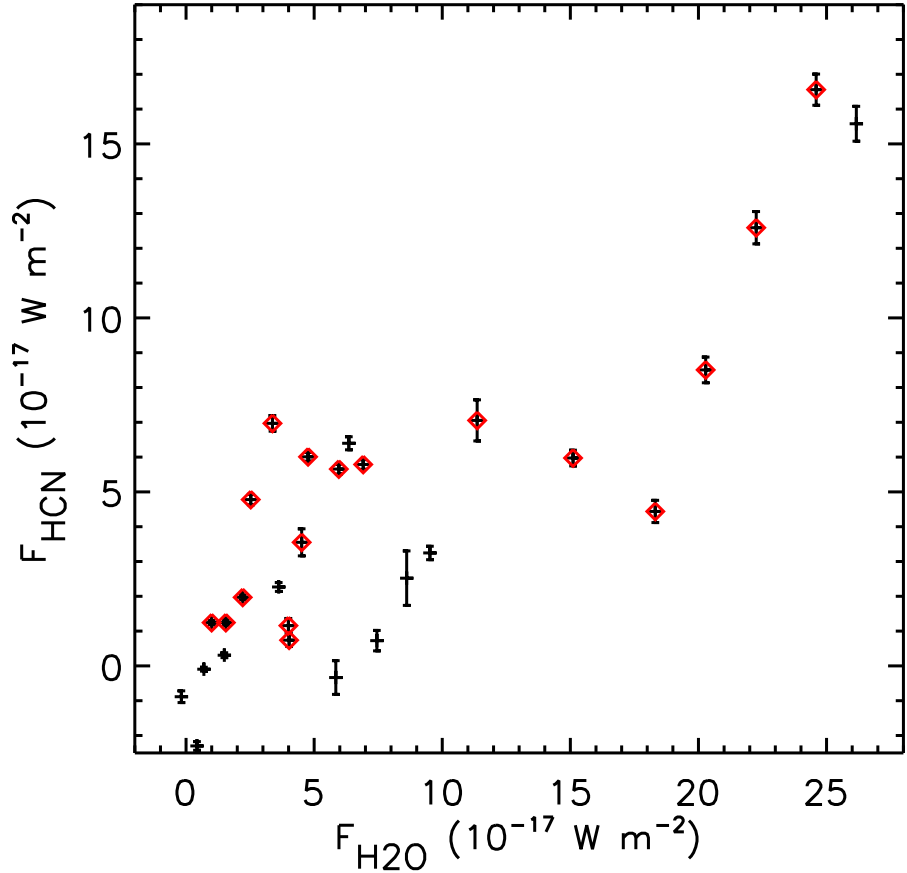}{6.0truein}{0}{100}{100}{-210}{70}
\caption{Emission strength of the $17\micron$ $\HtwoO$ emission 
features plotted against HCN emission strength. These quantities are 
well correlated, although there is significant dispersion, indicating 
the role of other parameters in determining the HCN emission strength.
Red diamonds indicate sources with measured disk masses that are typical 
T Tauri stars, i.e., they have the 
mid-infrared colors of normal T Tauri stars and do not have a strong jet.  }
\end{figure}

\begin{figure}
\plotfiddle{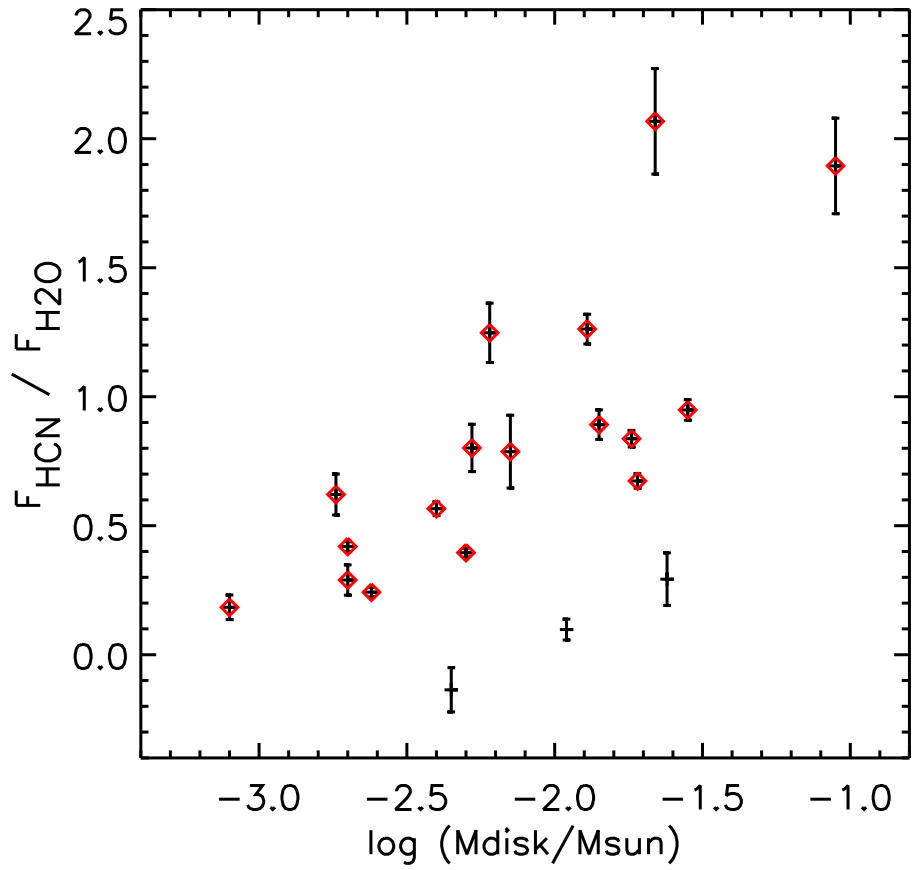}{6.0truein}{0}{100}{100}{-210}{70}
\caption{Ratio of the HCN and $\HtwoO$ emission strengths plotted 
against submillimeter disk mass.  
Red diamonds indicate sources with measured disk masses that are 
typical T Tauri stars, i.e., they have the 
mid-infrared colors of normal T Tauri stars and do not have 
a strong jet.  
}
\end{figure}

\begin{figure}
\plotfiddle{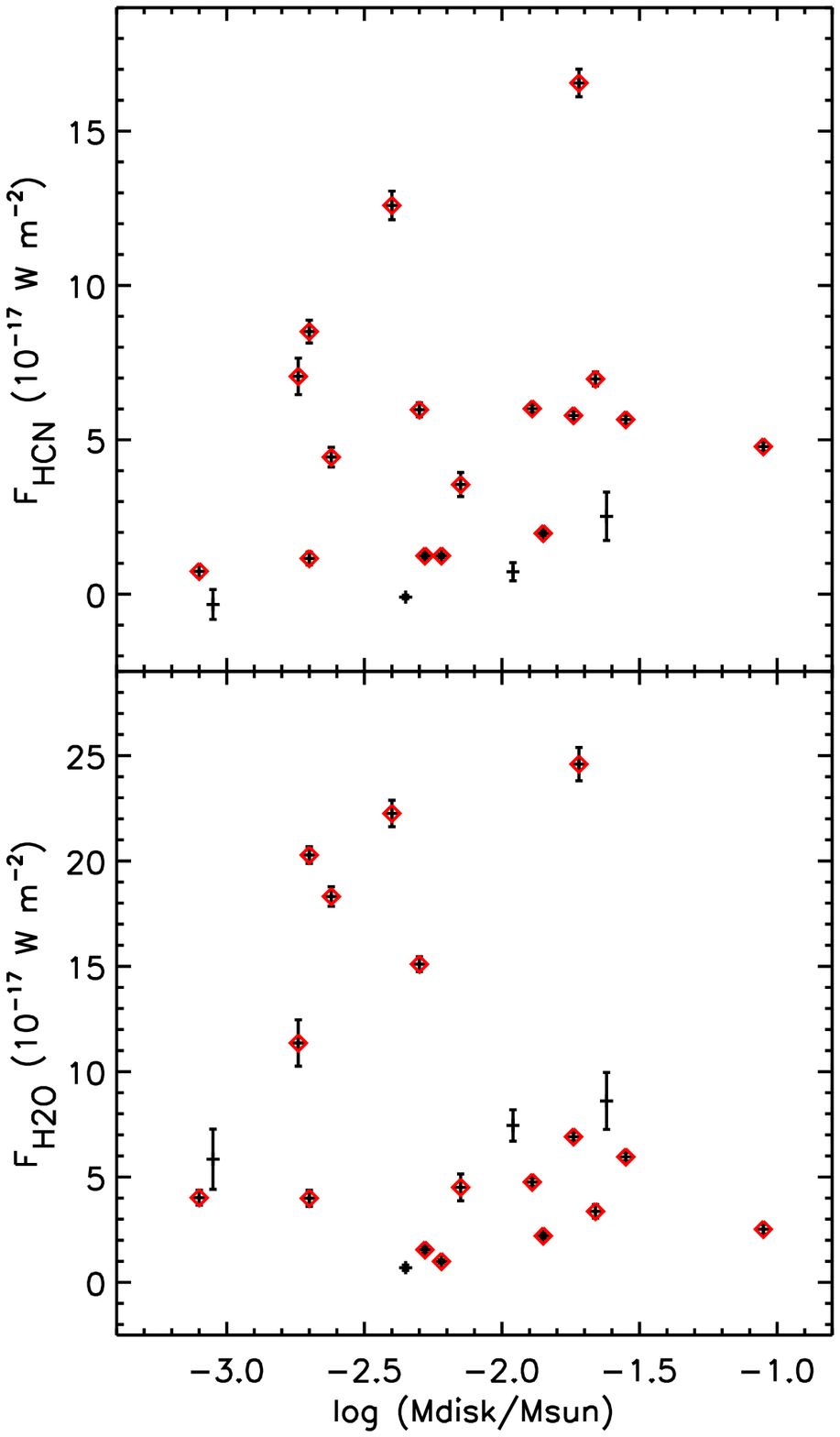}{6.5truein}{0}{90}{90}{-170}{-10}
\caption{HCN (top) and $\HtwoO$ (bottom) emission strengths plotted 
against submillimeter disk mass.  
Red diamonds have the same meaning as in Figure 3. 
No trends are apparent, in contrast with Figure 3. 
}
\end{figure}

\references

\noindent Ag\'undez, M., Cernicharo, J., \& Goicoechea, J.\ R.\ 2008, A\&A, 483, 831

\noindent Andrews, S.\ M., \& Williams, J.\ P. 2005, ApJ, 631, 1134

\noindent Carr, J.\ S.\ \& Najita, J.\ R.\ 2011, ApJ, 733, 102

\noindent Carr, J.\ S.\ \& Najita, J.\ R.\ 2008, Science, 319, 1504

\noindent Chiang, E., \& Youdin, A.\ 2010, ARAA, AREPS, 38, 493

\noindent Ciesla, F.\ J., \& Cuzzi, J.\ N.\ 2006, Icarus, 181, 178

\noindent Dullemond, K.\ \& Dominik, C.\ 2005, A\&A, 434, 971

\noindent Evans et al.\ 2003, PASP, 115, 965

\noindent Furlan, E., et al. 2006, ApJS, 165, 568

\noindent Greaves, J.\ S., Fischer, D.\ A., Wyatt, M.\ C., 
Beichman, C.\ A., \& Bryden, G.\ 2007, MNRAS 378, L1 

\noindent Hartigan, P., Strom, K.\ M., \& Strom, S.\ E.\ 1994, ApJ, 427, 961

\noindent Heinzeller, D., Nomura, H., Walsh, C., \& Millar, T. J.\ 2011, ApJ, 731, 115

\noindent Hillenbrand, L.\ A.\ 2008, Physica Scripta, 130, 014024

\noindent Houck, J.\ R.\ et al.\ 2004, ApJS, 154, 18

\noindent Itoh, et al.\ 2008, PASJ, 60, 209 

\noindent McCabe, C., Duch\^ene, G., \& Ghez, A.\ M.\ 2003, ApJ, 588, L113

\noindent Meijerink, R., Pontoppidan, K.\ M., Blake, G.\ A., Poelman, D.\ R., 
\& Dullemond, C.\ P.\ 2009, ApJ, 704, 1471

\noindent Najita, J.\ R., \'Ad\'amkovics, M., \& Glassgold, A.\ E.\ 2011, 
ApJ, 743, 147

\noindent \"Oberg, K.\ I.\ et al.\ 2011a, ApJ, 734, 98

\noindent \"Oberg, K.\ I., Murray-Clay, R., \& Bergin, E.\ A.\ 2011b, 
ApJ, 743, L16

\noindent Pascucci, I., Apai, D., Luhman, K., Henning, Th., Bouwman, J., 
Meyer, M.\ R., Lahuis, F., \& Natta, A.\ 2009, ApJ, 696, 143

\noindent Perez, L.\ M.\ et al.\ 2012, ApJ, 760, L17

\noindent Pontoppidan, K.\ M., Salyk C., Blake, G.\ A., 
Meijerink, R., Carr, J.\ S., \& Najita, J.\ 2010, ApJ, 720, 887

\noindent Salyk, C., Pontoppidan, K.\ M., Blake, G.\ A., Najita, J.\ R., 
\& Carr, J.\ S.\ 2011, ApJ, 731, 130

\noindent Salyk, C., Pontoppidan, K.\ M., Blake, G.\ A., Lahuis, F., 
van Dishoeck, E.\ F., \& Evans, N.\ J.\ II 2008, ApJ, 676, L49

\noindent Skemer, A.\ J., Close, L.\ M., Greene, T.\ P., Hinz, P.\ M., 
Hoffmann, W.\ F., \& Males, J.\ R.\ 2011, ApJ, 740, 43

\noindent Smith, I.\ W.\ M., \& Warr, J.\ F.\ 1991, J.\ Chem.\ Soc.\ Far.\ Trans.\ 87, 807

\noindent White, R.\ J.\ \& Ghez, A.\ M.\ 2001, ApJ, 556, 265

\noindent Willacy, K.\ \& Woods, M.\ 2009, ApJ, 703, 479

\noindent Woitke, P., Kamp, I., \& Thi, W.-F.\ 2009, A\&A, 501, 383

\noindent Wyatt, M.\ C., Clarke, C.\ J., \& Greaves, J.\ S.\ 2007, 
MNRAS, 380, 1737

\end{document}